# Metal-Free Room-Temperature Ferromagnetism

Hongde Yu[1], Thomas Heine[1, 2, 3]*

1 Faculty of Chemistry and Food Chemistry, TU Dresden, Bergstrasse 66c, 01069 Dresden, Germany.
2 Helmholtz-Zentrum Dresden-Rossendorf, Centrum for Advanced Systems Understanding, CASUS, Untermarkt 20, 02826 Görlitz, Germany.
3 Department of Chemistry, Yonsei University and IBS center for nanomedicine, Seodaemun-gu, Seoul 120-749, Republic of Korea.

**Abstract:**

Achieving robust room-temperature ferromagnetism in purely organic 2D crystals remains a fundamental challenge, primarily due to antiferromagnetic (AFM) coupling mediated by π-electron superexchange. Here, we present a mix-topology design strategy to induce strong ferromagnetic (FM) coupling in metal-free 2D systems. By covalently connecting radical polyaromatic hydrocarbon monomers (also referred to as nanographenes) with distinct sublattice topologies, this approach rationally breaks inversion symmetry and enables selective alignment of majority spins across the extended network, giving rise to metal-free ferromagnetism. Based on this strategy, we designed a family of 32 organic 2D crystals featuring spin-1/2 and mixed spin-1/2-spin-1 honeycomb lattices. Systematic first-principles calculations reveal that these materials are robust FM semiconductors with tunable spin-dependent bandgaps ranging from 0.9 to 3.8 eV. Notably, we demonstrate record-high magnetic coupling of up to 127 meV, spin-splitting energies exceeding 2 eV, and Curie temperatures surpassing 550 K, indicating thermal stability well above room temperature. The microscopic origin of the strong FM exchange stems from enhanced spin-orbital overlap and dominant direct exchange, while AFM superexchange is effectively suppressed. Our findings establish a generalizable design principle for realizing robust metal-free FM semiconductors and open new avenues for developing flexible and biocompatible magnets for next-generation spintronic and quantum technologies.

**Introduction:**

Metal-free magnetic materials are attracting increasing interest as candidates for next-generation spintronic, quantum information, and flexible electronic devices, owing to their lightweight character, environmental sustainability, and outstanding chemical tunability.[1–3] In contrast to conventional inorganic magnets, where magnetism arises from localized *d*- or *f*- orbitals of metal atoms, metal-free magnetism emerges from strong electron correlation within partially filled and delocalized π-orbitals.[2,4] This unique form of π-electron magnetism has been observed in a range of systems, including magic-angle twisted bilayer graphene,[5,6] graphene nanoribbons,[3,7,8] and low-dimensional polymers,[2,9–11] which host a variety of exotic quantum phases such as fractional quantum Hall effect,[12,13] unconventional superconductivity,[14–16] and Haldane phase.[2] Among these materials, organic two-dimensional (2D) crystals (O2DCs)—formed by covalently linking molecular building blocks into well-defined topologies, including monolayer 2D polymers and layer-stacked covalent organic frameworks (COFs)—have emerged as versatile platforms for engineering correlated π-electron magnetism.[1,4] Most O2DCs are made of closed-shell molecules and therefore diamagnetic. Combining open-shell monomers may preserve the spin polarization of the monomers and predominantly results in antiferromagnetic (AFM) ordering.[2,11] This is largely attributed to strong π-electron coupling and dominant superexchange interactions that favor antiparallel spin alignment.[1,17–19] In contrast, organic ferromagnetic (FM) materials with net magnetization and lifted Kramers' degeneracy are highly desirable for enabling electrical and optical control of spin degrees of freedom.[3] Achieving robust room-temperature ferromagnetism (RTFM) in purely organic systems would mark a major milestone,[3,5,6] paving the way for a new generation of flexible and biocompatible magnets with inherently low magnetic damping. Yet, despite significant theoretical and experimental efforts, the realization of RTFM in organic 2D crystals remains a formidable challenge.[20–23]

Overcoming this challenge requires not only the development of π-conjugated radicals as spin units, but also effective strategies to control magnetic interactions across π-extended frameworks.[4,11,24] Recent progress in constructing low-dimensional magnetic architectures—particularly those based on non-Kekulé polycyclic aromatic hydrocarbon (PAH) radicals, such as triangulenes[25,26] and Clar's goblet,[27,28] has been enabled by advances in on-surface synthesis. These systems have facilitated the exploration of correlated spin phenomena, including spinon excitations and Haldane phase with fractional edge states in 1D spin chains, and Mott-Hubbard insulating behavior in magnetic 2D networks.[2,11,29–34] However, in these π-conjugated systems, through-bond magnetic interactions between spin centers are predominantly AFM.[2,11,29,30] This AFM character originates from balanced sublattice topologies inherent to these systems, as explained by Lieb's theorem[35] (also known as Ovchinnikov rule[36]), which predicts a spin-degenerate, zero-magnetization ground state in half-filled bipartite lattices with inversion and time-reversal symmetries.[37,38]

Significant efforts have been devoted to breaking sublattice symmetry and inducing net magnetization in extended π-conjugated systems, such as graphene nanoribbons and spin chains.[3,24,39–41] For instance, Janus graphene nanoribbons bearing a single zigzag edge exhibit FM ground states, arising from sublattice asymmetry.[3] However, such sublattice imbalance in Lieb's theorem alone cannot guarantee to yield strong FM interactions. For example, weak FM couplings of approximately 1 meV and 7 meV have been observed in spin chains employing meta-phenylene bridges[39,40] or spin-sublattice engineering strategies,[24,41] although net magnetization is preserved according to Lieb's theorem. This can be attributed to the spatial separation of spin-orbitals and asymmetric connectivity suppressing electronic communication, leading to limited orbital overlap and spin-spin interaction. Therefore, the absence of strong through-bond FM coupling in π-conjugated, half-filled radical systems remains a major bottleneck toward realizing room-temperature, metal-free 2D ferromagnets. This persistent limitation underscores the urgent need for a generalizable design principle enabling strong, tunable FM coupling across extended 2D lattices.

Triangulene, the smallest triplet-ground-state nanographene, serves as a prototypical multi-radical unit for exploring metal-free magnetism in 2D honeycomb frameworks.[42] (Figure 1) Here, we introduce a mix-topology design strategy to realize strong FM coupling and robust RTFM in metal-free O2DCs. This approach leverages radical monomers with distinct sublattice topologies—denoted as Topology I and Topology II—to construct binary triangulene-based networks. (Figure 1) By rationally breaking the inversion symmetry, this approach allows for efficient overlap between half-filled π-orbitals, while enabling selective alignment of majority spins across the lattice. Using this strategy combined with first-principles calculations, we design a family of binary O2DCs and demonstrate that all these systems are FM semiconductors, some of them with unprecedented FM couplings and Curie temperatures well above room temperature. Due to the lifting of Kramers' spin degeneracy, these FM 2D systems demonstrate controllable spin-splitting as well as tunable spin-polarized bandgaps. This work establishes a general blueprint for metal-free, chemically tunable organic magnets and suggests new opportunities for flexible spintronic and quantum devices.

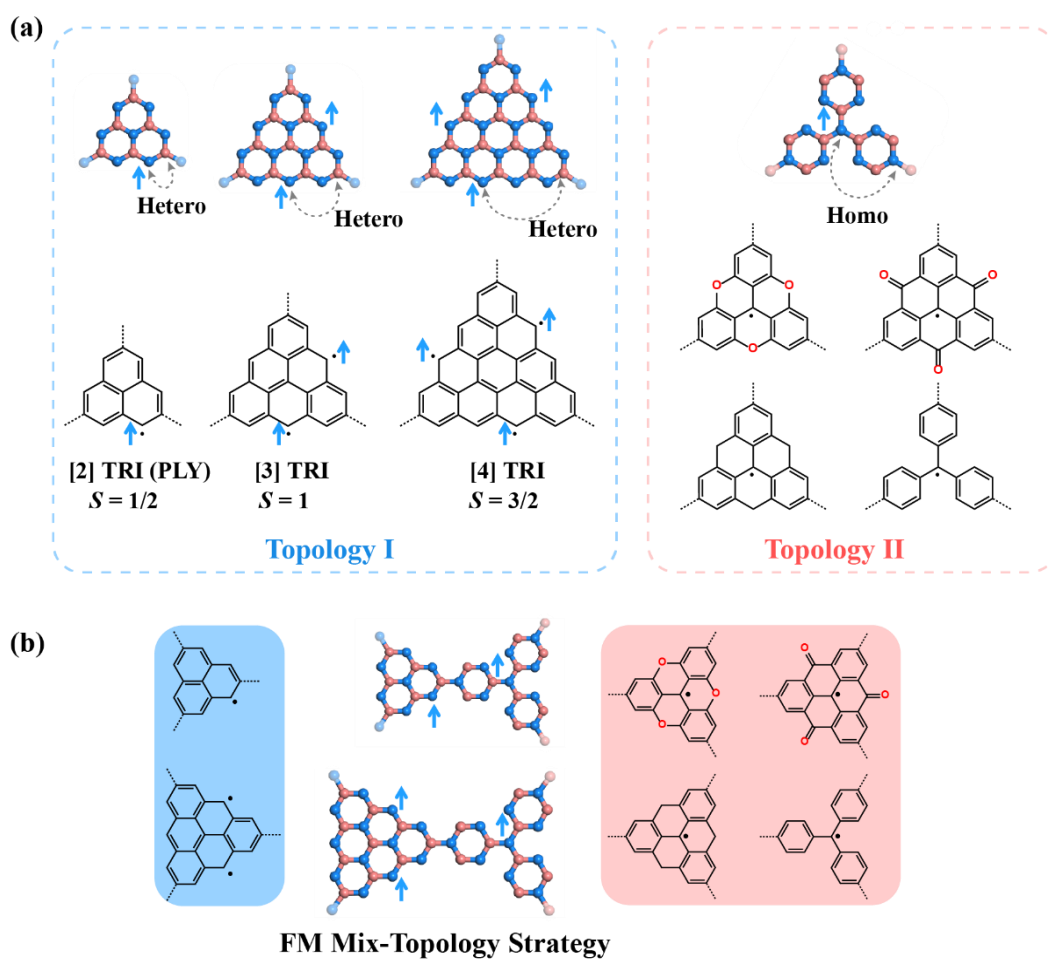

**Figure 1.** Mix-topology design strategy for achieving metal-free ferromagnetism in metal-free 2D systems with honeycomb lattice. Color coding: blue and red denote sublattices A and B, respectively, connectivity atoms of the next neighbors are marked in light blue or red, respectively. (a) Classification of radical monomers into Topology I and Topology II based on the distribution of spin density over the bipartite sublattices. Topology I monomers (left) feature majority spins on one sublattice and reaction-active connection sites on the opposite sublattice (hetero-spin configuration), while Topology II monomers (right) exhibit both majority spins and connecting sites on the same sublattice (homo-spin configuration). (b) Schematic representation of the mix-topology strategy, wherein Topology I and Topology II monomers are covalently linked to form binary organic 2D crystals (O2DCs) with FM spin alignment.

## Results and discussion:

### 1. Mix-topology design strategy for ferromagnetic organic 2D crystals

As shown in Figure 1a, triangulene (TRI)-based radical monomers can be categorized into two distinct topological classes—Topology I and Topology II—based on the sublattice identities of the majority spin sites and the reaction-active connection sites designated as bridging positions. In Topology I, the dominant spin-sublattice and the polymerization sublattice belong to opposite sublattices of the underlying honeycomb lattice. In contrast, Topology II monomers feature both the dominant spin density and the connecting sites on the same sublattice. (Figure 1a) A variety of triangular nanographenes with zigzag edges fall under Topology I, such as phenalenyl (PLY or [2] TRI),[45,46] [4]TRI,[47] and [5]TRI.[48] In these structures, both the total spin quantum number ($S$) and the number of energy-degenerate singly occupied molecular orbitals (SOMOs), corresponding to the zero-mode number, increase linearly with edge extension.[49] This behavior is consistent with Lieb's theorem, as one sublattice becomes dominant and increases the imbalance.[4,26,50,51] In contrast, many hetero-triangulenes exemplify Topology II, such as trioxatriangulene (TAM),[52] trioxotriangulene (TOT),[53] trihydrotriangulene (TRIH),[54] and triphenylmethyl (TPM).[55] These radicals have been synthesized via both solution-phase and on-surface methods.[45,46] They generally exhibit $D_{3h}$ symmetry with delocalized spin density across a planar, fully π-conjugated carbon skeleton, with the exception of TPM, which adopts a non-planar geometry with $C_3$ symmetry due to the steric hindrance between adjacent hydrogen atoms on the peripheral benzene rings.[56] (Figures 1a and S1)

When constructing complex π-conjugated magnetic architectures—such as spin dimers,[57,58] quantum rings,[59] and 1D or 2D polymers[2,4,29]—by covalently linking radical units, two distinct topological scenarios can arise: the connected spin units may share the same sublattice topology, or they may possess different ones. In the case where the spin units have the same Topology, the resulting magnetic interactions are typically AFM, consistent with the predictions of Lieb's theorem, leading to low-spin ground states. (Figures 1b and S2-4) Such AFM coupling has been widely observed in homogenous O2DCs. For example, [TRI] 2D networks show Topology I and exhibit AFM coupling of approximately -32 meV,[21] while TAM-based COFs, which are of

Topology II, show even stronger AFM exchange, up to -121 meV.[31] These systems often display Mott-Hubbard insulating behavior due to strong π-electron correlation. Interestingly, binary O2DCs composed of spin units with the same topology can also exhibit ferrimagnetic behavior.[37] A notable example is the [TRI-PLY] system, in which alternating spin-1 (TRI) and spin-1/2 (PLY) units are coupled antiferromagnetically. (Figures S2-S3) Due to the unequal spin magnitudes, the magnetic moments are only partially compensated, resulting in a net magnetization.[37] (Figures S2-S4)

Here, we introduce the mix-topology design strategy to realize through-bond FM coupling in 2D π-conjugated frameworks. (Figures 1b and 2a) This approach involves covalently linking radical building blocks with complementary sublattice topologies to construct extended 2D spin networks (specifically Topology I–Topology II). In such configurations, the unpaired spins on adjacent monomers align parallel across the lattice. According to Lieb's theorem, they favor high-spin ground states and thus enable intrinsic metal-free ferromagnetism. Leveraging this mix-topology concept, we computationally designed a series of 32 binary O2DCs, assembled from stable (hetero)triangulene-based radical monomers. The spin units are connected via either direct bonding or π-conjugated linear linkers, including ethynylene (–C≡C–, CC), butadiyne (–C≡C–C≡C–, CCCC), and para-phenylene (Ph) groups (Figures 2a and S5-12), facilitating delocalized spin density across the lattice.

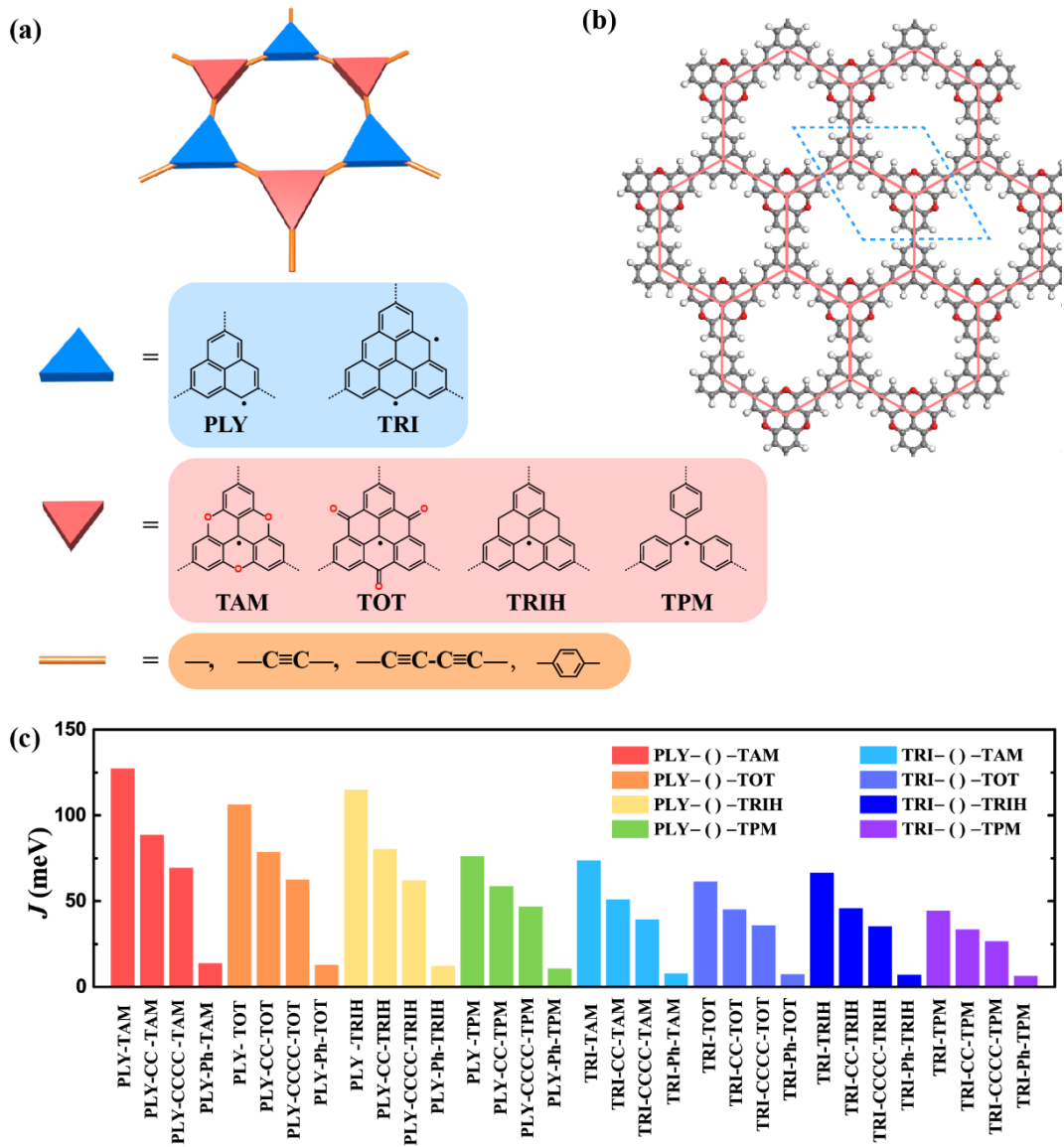

**Figure 2.** Structural design and magnetic coupling in FM organic 2D crystals (O2DCs) based on the mix-topology strategy. (a) Schematic illustration of the mix-topology approach, in which Topology I monomers (PLY and TRI) are combined with Topology II monomers (TAM, TOT, TRIH, and TPM) via conjugated linkers—acetylene, diacetylene, or para-phenylene—to construct extended 2D networks. (b) Atomic structure of a representative binary O2DC, [PLY-TAM], showing the periodic arrangement of Topology I and Topology II units in a honeycomb lattice geometry. (c) Predicted magnetic couplings ($J$) for 32 FM O2DC, obtained using hybrid density functional theory (PBE0 level).

## 2. Magnetic couplings in ferromagnetic organic 2D crystals

As illustrated in Figure 2b, these FM O2DCs adopt a honeycomb lattice reminiscent of graphene and hexagonal boron nitride (h-BN). (Figure S13) The half-filled, degenerate π-orbitals located on each radical unit introduce strong electron-electron correlations, driving spontaneous spin polarization. To validate the effectiveness of the mix-topology design strategy, we systematically investigated the magnetic properties of all 32 binary O2DCs using first-principles calculations. (see Methods for computational details) For all systems examined, the FM configuration was consistently found to be the energetically favorable one, surpassing both the AFM (or ferrimagnetic) and closed-shell singlet states. (Table 1 and Figure 3). The magnetic couplings ($J$) between neighboring spin centers are extracted using the Heisenberg-Dirac-van Vleck (HDVV) Hamiltonian $\hat{H} = -\sum_{<i,j>} J \hat{S}_i \hat{S}_j$. As shown in Table 1 and Figure 2c, the $J$ values span a wide range, from 44 meV to 127 meV. Among these, [PLY-TAM] exhibits the strongest FM coupling, with a $J$ value of 127 meV, representing the record-high through-bond FM interaction in purely organic 2D systems.[20,21,23] (Table 1 and Figure 2c) Other PLY-based systems, including [PLY-TOT], [PLY-TRIH], and [PLY-TPM], also display exceptionally strong FM couplings of 106 meV, 114 meV, and 76 meV, respectively—substantially exceeding those of previously reported organic and inorganic 2D magnetic materials.[3,24,39–41,63–65] While PLY-based O2DCs form uniform spin-1/2 honeycomb networks, TRI-based systems give rise to hybrid spin lattices composed of alternating spin-1 (TRI) and spin-1/2 (Topology II) units. This mixed spin-1/2-spin-1 2D architecture features enhanced quantum fluctuation, offering an ideal platform for exploring complex magnetic excited states. We also observed a linker-dependent trend in magnetic couplings. As the π-conjugated linker length increases—from direct coupling to ethynylene (–C≡C–), butadiyne (–C≡C–C≡C–), and para-phenylene (–Ph–)—the $J$ values decrease due to reduced spatial overlap between spin-orbitals. (Figures 2c and S14-S37)

**Table 1**. Magnetic and electronic properties of representative FM O2DCs constructed via the mix-topology strategy. The magnetic coupling constant ($J$), spin quantum number ($S$) of spin units, total magnetic moment ($M$) per unit cell, spin-splitting energy ($E_{split}$), spin-polarized bandgaps for spin-up and spin-down channels ($E_g^{\uparrow\uparrow}$ and $E_g^{\downarrow\downarrow}$), spin-flip gaps ($E_g^{\uparrow\downarrow}$) and Curie temperatures ($T_c$) are shown. To provide molecular insights, overlap integral ($S_o$), chemical potential (on-site energy) offset ($\Delta\epsilon$), and SOMO-to-LUMO energy gap ($\Delta E_{S-L}$) of the building units are included. All results are calculated at the PBE0 level. See Methods section for details.

|  | $J$ (meV) | $S$ | $M$ ($\mu_B$) | $E_{split}$ (eV) | $E_g^{\uparrow\uparrow}$ (eV) | $E_g^{\downarrow\downarrow}$ (eV) | $E_g^{\uparrow\downarrow}$ (eV) | $T_c$ (K) | $S_o$ | $\Delta\epsilon$ (eV) | $\Delta E_{S-L}$ (eV) |
|---|---|---|---|---|---|---|---|---|---|---|---|
| [PLY-TAM] | 127 | 1/2, 1/2 | 2 | 2.11 | 2.56 | 3.81 | 1.70 | 556 | 0.23 | 0.76 | 1.56 |
| [PLY-TOT] | 106 | 1/2, 1/2 | 2 | 1.31 | 2.67 | 2.27 | 0.96 | 470 | 0.27 | 1.36 | 0.59 |
| [PLY-TRIH] | 115 | 1/2, 1/2 | 2 | 1.64 | 3.19 | 3.63 | 1.98 | 506 | 0.21 | 0.48 | 1.85 |
| [PLY-TPM] | 76 | 1/2, 1/2 | 2 | 1.60 | 3.56 | 3.81 | 2.21 | 336 | 0.18 | 0.07 | 2.26 |
| [TRI-TAM] | 74 | 1, 1/2 | 3 | 1.99 | 2.51 | 3.75 | 1.75 | 570 | 0.18 | 0.88 | 1.58 |
| [TRI-TOT] | 61 | 1, 1/2 | 3 | 1.15 | 2.64 | 2.08 | 0.93 | 470 | 0.24 | 1.25 | 0.69 |
| [TRI-TRIH] | 66 | 1, 1/2 | 3 | 1.55 | 3.09 | 3.54 | 1.99 | 508 | 0.18 | 0.58 | 1.88 |
| [TRI-TPM] | 44 | 1, 1/2 | 3 | 1.45 | 3.45 | 3.72 | 2.27 | 338 | 0.21 | 0.17 | 2.29 |

The Lieb theorem—derived from the standard Hubbard model for half-filled bipartite lattices—offers qualitative insight into the emergence of magnetization in mix-topology O2DC. In order to quantitatively rationalize these strong FM couplings, an extended Hubbard Hamiltonian incorporating both chemical potential asymmetry and direct magnetic exchange is employed:

$$H = -t \sum_{\langle i,j \rangle, \sigma} \left( c_{i,\sigma}^+ c_{j,\sigma} + c_{j,\sigma}^+ c_{i,\sigma} \right) + U \sum_i n_{i\uparrow} n_{i\downarrow} + \sum_i \epsilon_i \, n_i + K \sum_{\langle i,j \rangle} \sum_{\sigma, \sigma'} c_{i,\sigma}^+ c_{j,\sigma'}^+ c_{i,\sigma'} c_{j,\sigma}$$

Here, $t$ and $U$ denote the inter-site electronic hopping integral (i.e., electronic coupling) and on-site Coulomb repulsion, respectively, as in the conventional Hubbard model. The additional terms $\epsilon$ and $K$ represent the on-site energy (i.e., chemical potential) and direct magnetic exchange, respectively. The $\epsilon$ term—analogous to the chemical potential in tight-binding models—is particularly relevant in binary lattices such as mix-topology O2DCs and h-BN. The interplay between $\Delta\epsilon$ and $U$ governs the emergence of diverse quantum phases: a band insulator arises when $\Delta\epsilon \gg U$, as in h-BN, while a Mott-Hubbard insulator is favored when $U \gg \Delta\epsilon$, as seen in homo-topological AFM π-conjugated frameworks. In the mix-topology 2D O2DCs studied here, the on-site energy (chemical potential) offset $\Delta\epsilon$ —arising from energy mismatches between the SOMOs of the two monomers—ranges from 0.1 to 1.4 eV. (Table 1) For instance, in [PLY-TAM], $\Delta\epsilon$ is 0.76 eV, substantially smaller than the on-site Coulomb repulsion $U$ of 2.66 eV for the PLY unit,[21] placing these systems firmly within the Mott-insulating regime with inherent spin-polarization. This energy mismatch, together with intrinsically uncompensated sublattice, significantly suppresses the hopping integral $t$, such as ~ 1 meV in [PLY-TAM]. (Figure S13) In the strongly correlated limit, $U \gg t$, the overall magnetic interaction $J$ is determined by the competition between FM direct exchange $K$ and AFM kinetic superexchange $-4t^2/U$, following: $J = K - 4t^2/U$, as prescribed by the Goodenough-Kanamori rule.[66] In mix-topology 2D O2DCs, the dominance of direct exchange arises from the extremely small $t$, which renders the kinetic exchange negligible. However, the orbital overlap integrals ($S_o$) between neighboring radical units are substantial. For example, [PLY-TAM] and [PLY-TOT] exhibit $S_o$ values of 0.23 and 0.27, respectively, indicating strong spin-orbital overlap and spin-spin interactions. (Table 1) These values are comparable to those found in strongly AFM-coupled systems such as PLY-PLY ($J$ = -45 meV, $S_o$ = 0.13).[67] Compared to inorganic magnets with localized d-/f-orbitals, the spin-polarized π-orbitals in these conjugated nanographene frameworks are delocalized, enabling larger spatial overlap and enhanced direct exchange. This substantial overlap not only supports extended π-conjugation but also ensures efficient FM coupling across the 2D lattice, validating the core premise of the mix-topology design.

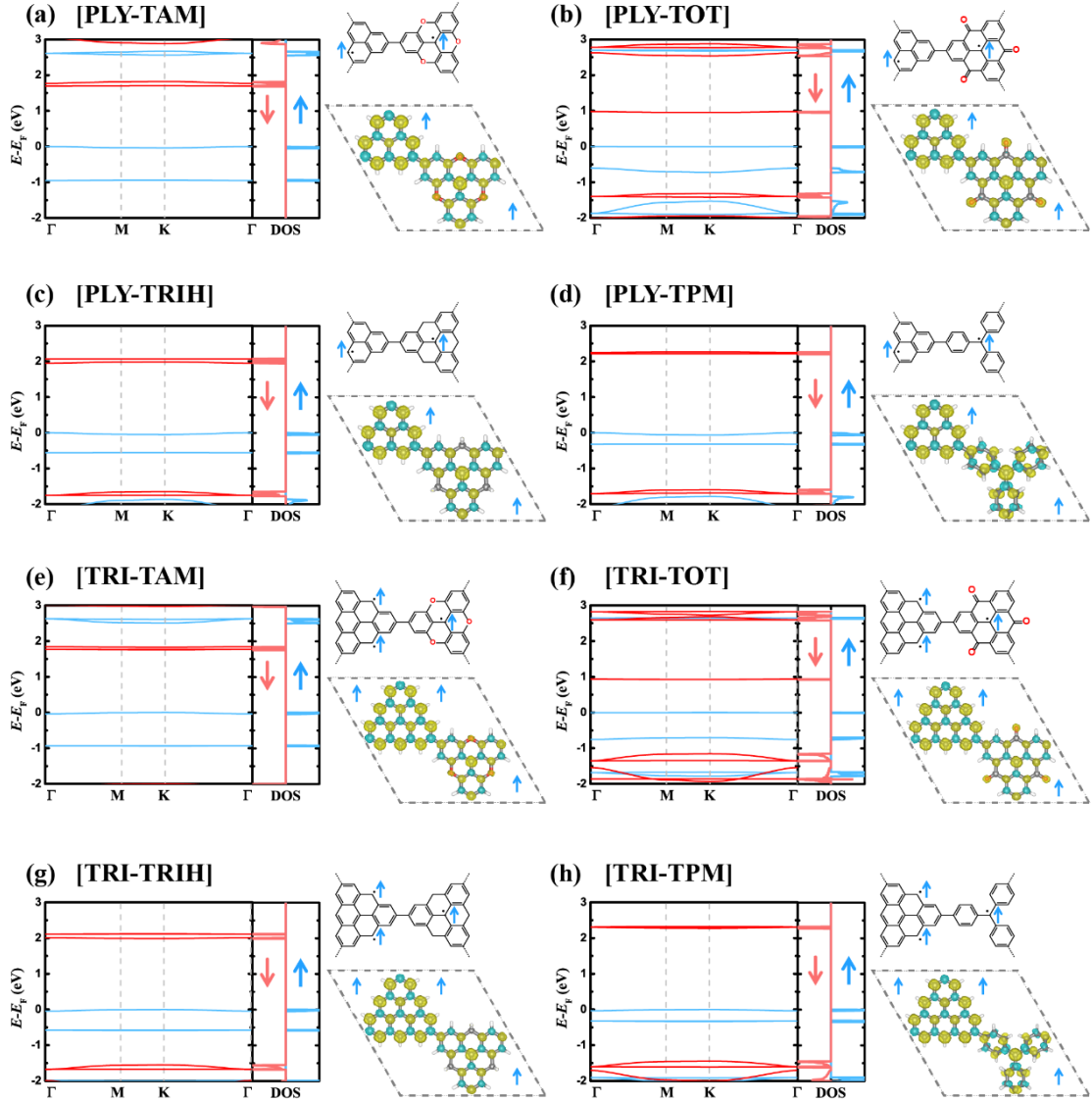

**Figure 3**. Spin-polarized band structures and spin densities of FM O2DCs. (a-h) Spin-polarized band structures, densities of states (DOS), molecular structures, and spin density distributions for [PLY-TAM], [PLY-TOT], [PLY-TRIH], [PLY-TPM], [TRI-TAM], [TRI-TOT], [TRI-TRIH], and [TRI-TPM], respectively. Blue and red lines denote spin-up and spin-down channels.

## 3. Spin-polarized electronic structures of FM O2DCs

We further demonstrate that the designed O2DCs are FM semiconductors featuring spin-polarized flat bands and spin-dependent bandgaps. (Figures 3 and S14-S37) These flat bands, located near the Fermi level, indicate electronic states that are energetically localized yet delocalized in momentum space. Unlike atomic flat bands—typically associated with dangling bonds without inter-site interaction[68]—these π-electron flat bands arise from correlated molecular orbitals with substantial wavefunction overlap. Within such bands, many-body interactions dominate over kinetic energy, giving rise to a high density of states at the Fermi level and enabling strong correlation effects. The coexistence of flat bands and strong orbital overlap creates an ideal platform for realizing strongly correlated quantum phases, such as fractional quantum Hall states and unconventional superconductivity, akin to phenomena observed in magic-angle twisted bilayer graphene and O2DCs with kagome lattice.[14,69,70] Similar spin-polarized flat bands have also been identified in AFM-coupled O2DCs, such as [TRI], [PLY], and [TAM].[20,21,31] In these Mott-Hubbard AFM insulators, however, due to the preservation of inversion and time-reversal symmetries, the spin-up and spin-down channels remain energetically degenerate, in agreement with Kramers' theorem.[31,38] In sharp contrast, all the designed FM O2DCs exhibit pronounced spin-splitting, arising from the breaking of inversion symmetry and thus lifting Kramers' spin degeneracy.[71] (Table 1, Figures 3 and 4d) In these FM systems, the spin-splitting energies range from 1.15 eV to 2.11 eV. These large spin-splittings are unprecedented among reported magnetic 2D polymers,[31] offering exciting opportunities for spintronic applications, such as organic spin valves and spin-selective transport devices, where carrier transport can be controlled via modulation of Fermi level.

In addition to large spin-splitting, these FM O2DCs feature tunable spin-dependent bandgaps, including both spin-flip and spin-polarized (spin-conserving) transitions (Table 1, Figure 3, and Figures 4e-g). The spin-flip bandgap ($E_g^{\uparrow\downarrow}$) is defined as the lowest energy excitation between states of opposite spin channels. This quantity is particularly relevant to charge transport under electrical gating, where carriers of both spin species are accessible via modulating the Fermi energy. By contrast, spin-polarized bandgaps ($E_g^{\uparrow\uparrow}$ and $E_g^{\downarrow\downarrow}$) refer to the minimal excitation energies within a single spin channel, corresponding to spin-conserving optical transitions. These gaps determine the intrinsic optical response of spin-polarized carriers and are critical for opto-spintronic applications. As shown in Figure 3 and Table 1, the spin-flip bandgaps across the binary FM O2DCs span a broad range from 0.93 eV to 2.21 eV. Among these, [PLY-TPM] exhibits the largest $E_g^{\uparrow\downarrow}$ of 2.21 eV, while [TRI-TOT] presents the smallest value of 0.93 eV. Interestingly, for pairs of polymers containing the same Topology II monomer, the corresponding PLY-based and TRI-based systems exhibit comparable $E_g^{\uparrow\downarrow}$—for instance, 1.64 eV in [PLY-TRIH] and 1.55 eV in [TRI-TRIH]. (Table 1) This suggests that the Topology II unit predominantly governs the spin-flip bandgap, arising from the similar frontier orbital energies between the Topology I monomers (TRI and PLY). (Figure 4b and Table S1) Due to the staggered alignment of SOMO and lowest unoccupied molecular orbitals (LUMO) levels between the two monomers, the spin-flip transitions are dominated by the LUMO of the Topology II unit. This mechanism is further supported by a positive correlation between the SOMO-LUMO energy difference ($\Delta E_{S-L}$) of the monomers and $E_g^{\uparrow\downarrow}$, promising for controlling spin-flip bandgaps at the molecular level. (Figure 4c)

In comparison to the spin-flip bandgaps, the spin-polarized bandgaps—$E_g^{\uparrow\uparrow}$ and $E_g^{\downarrow\downarrow}$—are significantly larger, ranging from 2.0 to 3.8 eV. (Table 1, Figures 4f, 4g) Similarly, both $E_g^{\uparrow\uparrow}$ and $E_g^{\downarrow\downarrow}$ exhibit a strong dependence on the Topology II monomer. Among them, TOT-based polymers show relatively small spin-polarized bandgaps; for example, [PLY-TOT] exhibits $E_g^{\uparrow\uparrow}$ = 2.67 eV and $E_g^{\downarrow\downarrow}$ = 2.27 eV. These spin-dependent bandgaps enable spin control via external stimuli such as electrical gating or optical excitation. For instance, in [PLY-TAM], a gate voltage of ~1.8 V can shift the Fermi level to selectively activate the spin-down conduction channel, thereby facilitating spin-polarized transport. When combined with the inherently weak spin-orbit coupling and long spin coherence times in carbon-based materials, these FM O2DCs emerge as promising candidates for high-fidelity spintronic and quantum devices.

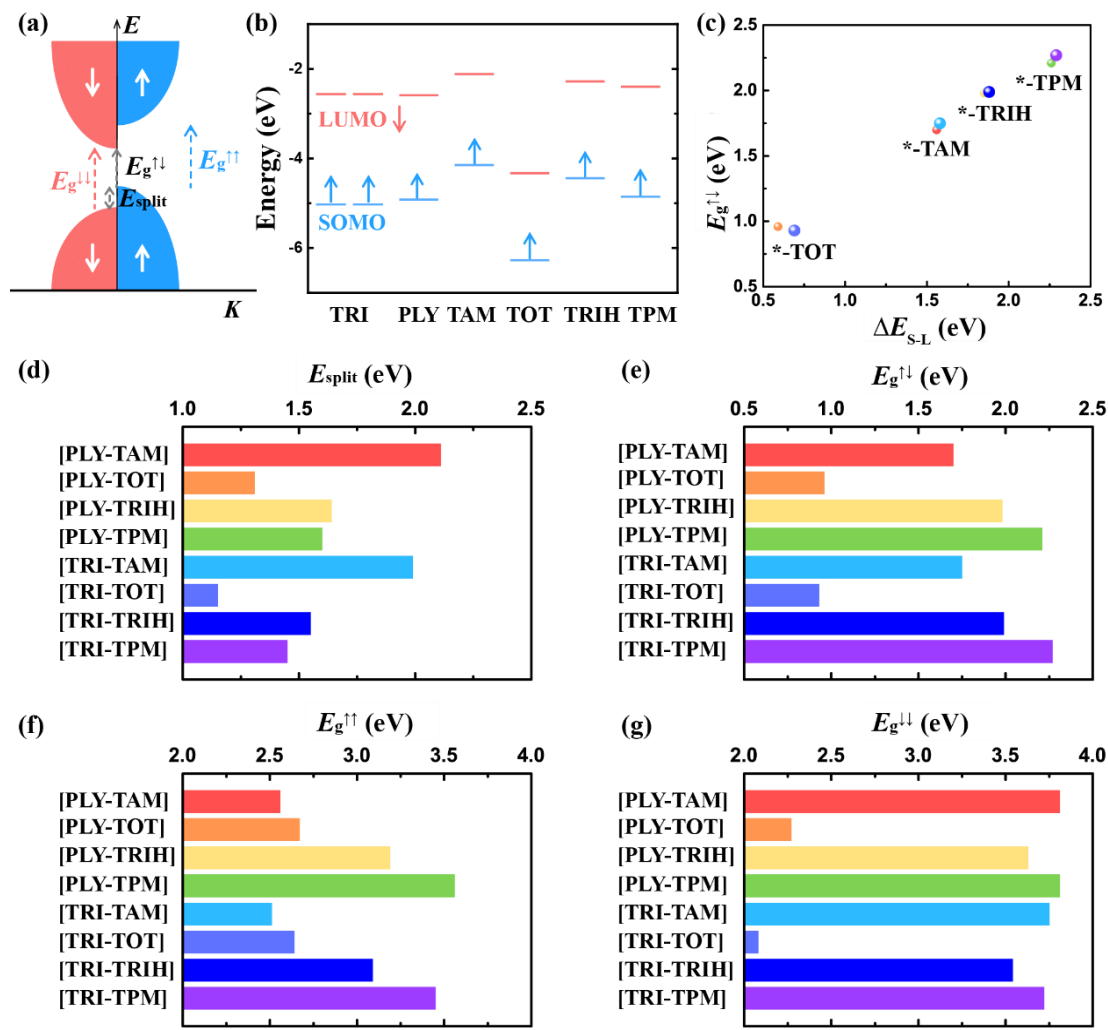

**Figure 4.** Electronic and magnetic properties of FM O2DCs. (a) Schematic illustration of spin-polarized features, illustrating the definitions of spin-splitting energy ($E_{split}$), spin-flip bandgaps ($E_g^{\uparrow\downarrow}$), and spin-polarized bandgaps for spin-up and spin-down channels ($E_g^{\uparrow\uparrow}$ and $E_g^{\downarrow\downarrow}$). (b) Frontier molecular orbital energies of radical monomers, showing singly occupied molecular orbitals (SOMOs, blue) and lowest unoccupied molecular orbitals (LUMOs, red). (c) Correlation between spin-flip bandgap ($E_g^{\uparrow\downarrow}$) and the SOMO-LUMO energy difference ($\Delta E_{S-L}$) across the series of FM O2DCs. (d-g) Predicted $E_{split}$, $E_g^{\uparrow\downarrow}$, $E_g^{\uparrow\uparrow}$ and $E_g^{\downarrow\downarrow}$ for selected FM O2DCs.

## 4. Thermal stability and feasibility of ferromagnetic ordering in organic 2D crystals

To evaluate the thermal robustness of the FM ordering in these O2DCs, we estimated their Curie temperatures ($T_c$) using Monte Carlo (MC) simulations. (Figures 5a and 5b). All systems exhibit spontaneous FM phase transitions well above room temperature, consistent with the large magnetic exchange constants obtained from first-principles calculations. Among these materials, [TRI-TAM] shows the highest critical temperature of 570 K, closely followed by [PLY-TAM] with a $T_c$ of 556 K, driven by its exceptionally strong nearest-neighbor coupling constants ($J = 127$ meV). In contrast, [PLY-TPM] shows the lowest $T_c$ of 336 K. (Figures 5b and S38-S39 and Table 1) The predicted $T_c$ values are unprecedented among metal-free systems, and notably exceed those of many known inorganic 2D van der Waals ferromagnets, such as $Cr_2Ge_2Te_6$ ($J \approx 4$ meV, $T_c \approx 66$ K) and $CrI_3$ ($J \approx 4$ meV, $T_c \approx 45$ K).[63,64] We acknowledge the long-standing theoretical debate regarding the existence of long-range FM order in low-dimensional systems, particularly in metal-free materials where spin-orbit coupling and magnetic anisotropy are minimal.[72] According to the Mermin-Wagner theorem, continuous symmetries preclude long-range magnetic order in infinite 2D systems at any finite temperature.[73] However, many experimental reports have demonstrated robust FM ordering in finite-sized 1D and 2D systems, such as cobalt atomic chains,[74] zigzag-edged graphene nanoribbons,[75] and Janus graphene nanoribbons.[3] Recent MC simulation studies have shown that, for isotropic 2D magnets with short-range interactions and weak spin-orbit coupling, long-range FM order can be stabilized in finite-sized systems at experimentally accessible scale—up to several micrometers—due to a substantial magnetic correlation length.[76] Specifically, the correlation length $\xi$ scales exponentially with the ratio $J/T$, i.e., $\xi \propto \exp(cJ/T)$, where c is a material-specific constant.[77] Therefore, for systems with exceptionally large $J$, as in our FM O2DCs, magnetic correlations can persist over mesoscopic or even macroscopic scales, enabling stable FM order in modern nanotechnology devices.

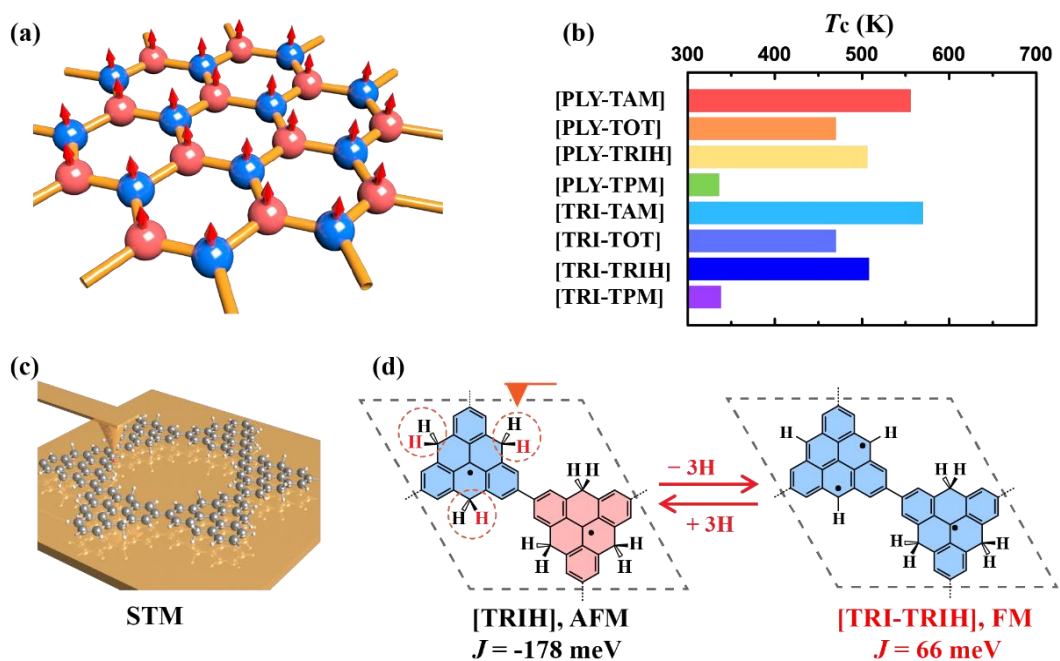

**Figure 5**. Ferromagnetic ordering and conceptual synthetic route for binary O2DCs. (a)

Schematic illustration of a metal-free 2D spin lattice with honeycomb topology, showing long-range FM order established through mix-topology design. (b) Estimated Curie temperatures ($T_c$) derived from Monte Carlo simulations, confirming thermal stability of the FM phase well above room temperature. (c) A conceptual framework for reversible modulation of magnetic interactions in O2DCs via selective hydrogenation and dehydrogenation, enabling controllable switching between AFM and FM states.

The proposed binary O2DCs can be synthesized via two strategies: (i) well-established Suzuki-Miyaura cross-coupling reaction between corresponding halide and organoboron precursors,[60] or (ii) tip-induced site-selective dehydrogenation of homogenous O2DCs using STM. (Figure 5c) The latter approach has been extensively demonstrated in previous experimental studies for a variety of π-radical monomers, including TRI, Clar's goblet, and Olympicene, as well as their corresponding polymers.[30,43,44] As a representative example, the fabrication of the [TRI-TRIH] binary O2DC can be achieved in a two-step process. (Figure 5d) First, Ullmann coupling, widely employed in on-surface synthesis, can be used to construct a homogenous 2D network of TRIH monomers, forming [TRIH], in which the radical centers are AFM coupled. Subsequently, by employing an STM tip to selectively dehydrogenate one TRIH unit per unit cell, a binary [TRI-TRIH] structure is formed, featuring alternating spin-1 (TRI) and spin-1/2 (TRIH) sites. As demonstrated earlier, this system exhibits strong FM coupling of 66 meV, giving rise to an ordered FM ground state with a predicted Curie temperature exceeding 500 K. Importantly, this site-selective dehydrogenation technique is not limited to 2D networks. It can also be extended to the fabrication of FM-coupled spin dimers and 1D polymers, enabling the exploration of low-dimensional magnetic phenomena such as spin waves and magnon excitations.

**Conclusion:**

In summary, we have established a mix-topology design strategy to achieve robust, room-temperature ferromagnetism in O2DCs. By covalently linking π-radical monomers with complementary sublattice topologies, we rationally break inversion symmetry and selectively align majority spins across the framework. This design suppresses AFM superexchange and enables strong through-bond FM coupling, driven by substantial spin-orbital overlap and direct exchange. Through systematic first-principles investigations of 32 binary O2DCs, we uncover a class of metal-free FM semiconductors exhibiting unprecedented magnetic and electronic properties, including magnetic coupling up to 127 meV, spin-splitting energies exceeding 2.1 eV, and Curie temperatures surpassing 550 K—all within fully conjugated, lightweight carbon-based frameworks. We further show that these systems support spin-polarized flat bands, tunable spin-dependent bandgaps, and gate-tunable spin transport, facilitating the generation of pure spin currents and opening avenues toward energy-efficient, scalable quantum devices. Moreover, we propose experimentally accessible fabrication routes—via Suzuki-Miyaura cross-coupling and STM-induced dehydrogenation—to realize these materials with atomic precision. Our findings establish a generalizable framework for designing room-temperature, metal-free ferromagnets, and chart a path forward for integrating lightweight and flexible quantum materials into next-generation spintronic, optoelectronic, and quantum information technologies.

**Acknowledgments**
This project has been funded by the Alexander von Humboldt Foundation and Deutsche Forschungsgemeinschaft within CRC 1415 and RTG 2861. The authors gratefully acknowledge the computing time provided to them on the high-performance computers Noctua 2 at the NHR Center PC2. These are funded by the Federal Ministry of Education and Research and the state governments participating on the basis of the resolutions of the GWK for the national high-performance computing at universities. The authors also thank the Center for Information Services and High-Performance Computing (ZIH) at TU Dresden for computational resources.

**Data Availability Statement**
The datasets generated in this work are available in the zenodo repository, under the accession code DOI: 10.5281/zenodo.15756816. All other relevant data supporting the findings of this study are available from the corresponding author upon reasonable request.

**Methods:**
Density functional theory (DFT) calculations:

The structural optimizations of all O2DCs were performed within the framework of density functional theory (DFT) using the Vienna Ab Initio Simulation Package (VASP 5.4.4).[61] A plane-wave cutoff energy of 400 eV was adopted, and the electron-ion interactions were described by the projector augmented wave (PAW) method.[78] The Perdew-Burke-Ernzerhof (PBE)[79] exchange-correlation functional within the generalized gradient approximation was employed, together with Grimme's D3 dispersion correction to account for van der Waals interactions.[80] The optimization criteria were set to a force threshold of 0.005 eV Å$^{-1}$, and an electronic energy convergence of $10^{-5}$ eV per self-consistent field cycle. A Monkhorst-Pack $k$-point grid of 8 × 8 × 1 was used for sampling the Brillouin zone during structural relaxations. Hybrid functional calculations were subsequently performed at the PBE0 level using the POB-TZVP basis set to obtain the band structure, density of states (DOS) and spin density, as implemented in the CRYSTAL17 software package,[62] with a k-point mesh of 16 × 16 × 1. Based on our prior benchmark study, the predicted magnetic couplings ($J$) of PBE0 functional closely align with high-level multi-reference calculations in metal-free systems.[81] Additionally, we provide Heyd–Scuseria–Ernzerhof (HSE06) results for comparison, which yields comparable $J$ values. (Table S2) Given that the systems comprise only C, H, and O elements, spin-orbit coupling (SOC) effects were neglected. $J$ values were evaluated based on the energy difference between the ferromagnetic (FM) and antiferromagnetic (AFM) states, following the expression $J = (E_{AFM}-E_{FM})/2zS^2$, where $z$ is the number of nearest neighbors (e.g., $z = 3$ for honeycomb lattices), and $S$ is the spin quantum number on each site, (e.g., $S = 1$ for TRI, $S = 1/2$ for PLY, TOT, TAM, TRIH and TPM). The electronic coupling parameter ($t$) was approximated from the bandwidth ($W$) using $W = 2zt$, extracted from the band structures of diamagnetic states. The on-site energy offset in Table 1 is calculated by the SOMO energy difference of radical building blocks. Overlap integrals between adjacent monomers were computed using dimer models within the unit cell via the Multiwfn program.[82] Frontier orbital energies (SOMO and LUMO) were obtained from PBE0/def2-TZVP single-point calculations using the Gaussian 16 software package.[83,84] All optimized geometries necessary to reproduce the results are available through the Zenodo repository.[85]

Monte Carlo (MC) simulations:
We have used MC simulations with the metropolis algorithm to determine the Curie temperature of all the FM O2DCs. The specific electronic heat capacity ($C_v$) has been calculated by $C_v = \frac{<E^2> - <E>^2}{k_B T^2}$. We have used a 40 × 40 × 1 supercell in the simulation and calculated 50 trajectories at every temperature. For each trajectory, we have calculated 100000000 steps while the last 90000000 steps are used to average and generate the data.